\documentclass[%
 reprint,
superscriptaddress,
 amsmath,amssymb,
 aps,
 prl
]{revtex4-2}

\usepackage{graphicx}
\usepackage{dcolumn}
\usepackage{bm}
\usepackage[colorlinks,citecolor=blue,urlcolor=blue,linkcolor=blue]{hyperref}

\begin{document}

\preprint{APS/123-QED}

\title{Flow rectification in loopy network models of bird lungs}

\author{Quynh M. Nguyen}
\affiliation{%
 Applied Math Lab, Courant Institute, 
 New York University, New York 10012, USA
}%

\affiliation{%
 Physics Department, New York University, New York 10003, USA}
 
 \author{Anand U. Oza}%
 \affiliation{%
 Department of Mathematical Sciences, New Jersey Institute of Technology, Newark, New Jersey 07102, USA
}%
 
\author{Joanna Abouezzi}%
 \affiliation{%
 Applied Math Lab, Courant Institute, 
 New York University, New York 10012, USA
}%

\author{Guanhua Sun}%
 \affiliation{%
 Applied Math Lab, Courant Institute, 
 New York University, New York 10012, USA
}%

\author{Stephen Childress}%
 \affiliation{%
 Applied Math Lab, Courant Institute, 
 New York University, New York 10012, USA
}%

\author{Christina Frederick}%
 \affiliation{%
 Department of Mathematical Sciences, New Jersey Institute of Technology, Newark, New Jersey 07102, USA
}%
 
 \author{Leif Ristroph}%
 \email{ristroph@cims.nyu.edu}
 
\affiliation{%
 Applied Math Lab, Courant Institute, 
 New York University, New York 10012, USA
}%



\date{\today}

\begin{abstract}

We demonstrate flow rectification, valveless pumping or AC-to-DC conversion in macroscale fluidic networks with loops. Inspired by the unique anatomy of bird lungs and the phenomenon of directed airflow throughout the respiration cycle, we hypothesize, test and validate that multi-loop networks exhibit persistent circulation or DC flows when subject to oscillatory or AC forcing at high Reynolds numbers. Experiments reveal that disproportionately stronger circulation is generated for higher frequencies and amplitudes of the imposed oscillations, and this nonlinear response is corroborated by numerical simulations. Visualizations show that flow separation and vortex shedding at network junctions serve the valving function of directing current with appropriate timing in the oscillation cycle. These findings suggest strategies for controlling inertial flows through network topology and junction connectivity.


\end{abstract}

\maketitle




Oscillatory, random or otherwise undirected movements can be induced into progressive motion by the presence of asymmetries. Rectification of fluids is a form of pumping \cite{glezer2002synthetic,prakash2008surface,lagubeau2011leidenfrost,thiria2015ratcheting,mo2016passive}, which is conventionally achieved via valves whose opening and closing motions are biased to direct flows appropriately. For example, the circulatory system involves pulsations produced by the beating heart that are rectified by flap-like valves to drive directed flow of blood through vessels. It is interesting and potentially useful that fluidic AC-to-DC conversion can also be achieved without moving elements but instead with entirely static geometries. This is permissible due to fluid dynamical irreversibility at high Reynolds numbers, for which the dominance of inertia over viscosity leads to phenomena such as flow separation and vortex shedding that respond sensitively to geometry and thus directionality \cite{tritton2012physical,schlichting2016boundary}. Such effects are exploited in fluidic diodes, devices whose asymmetric internal geometries lead to direction-dependent hydraulic resistance. Examples include conduits with diverging or converging walls, sawtoothed corrugations, and more complex geometries proposed by Nikola Tesla and others \cite{tesla1920valvular,groisman2004microfluidic,thiria2015ratcheting,tao2020microfluidic}.

\begin{figure*}
\centering
\includegraphics[width=16.5cm]{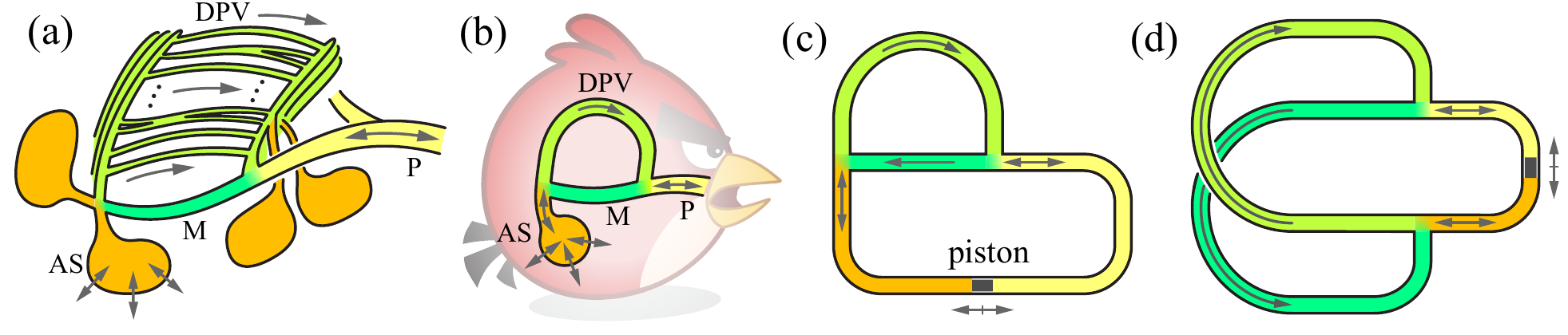}\vspace{-0.2cm}
\caption{Loopy network models of bird lungs. (a) Schematic of anatomy modified from \cite{scheid1972mechanisms}. Reciprocal expansion and contraction of air sacs (AS) drives oscillatory inhalation and exhalation flows in the primary bronchus (P). Directed flow is observed in dorso-para-ventro-bronchi (DPV) that span the mesobronchus (M). (b) A `spherical bird' simplification with a single AS and DPV. (c) A closed network formed by connecting P to AS, the latter replaced by a reciprocating piston. (d) A symmetrized network that modifies conduit lengths while preserving the connectivity of the T-junctions. }
\label{fig1}\vspace{-0.4cm}
\end{figure*}

Here we explore an alternative strategy for flow rectification in which the topology and connectivity of a network of symmetric pipes or channels leads to persistent circulation in response to oscillatory forcing. Our inspiration comes from a remarkable physiological observation: While air flow in the lungs of mammals oscillates with inhalation and exhalation, directional flows throughout the respiration cycle have been measured in birds \cite{hazelhoff1951structure,cohn1968respiration,bretz1971bird,brackenbury1971airflow,scheid1972mechanisms,powell1981airflow,maina2017pivotal,jones1981control}. This flow pattern may contribute to respiratory efficiency and has been viewed as an adaptation for the high metabolic demands of flight, although recent discoveries in reptiles suggest that the trait is ancestral \cite{farmer2010unidirectional,schachner2014unidirectional,cieri2016unidirectional}. The bird lung differs fundamentally from that of mammals in that its conduits are not only branched but also \textit{reconnect} to form \textit{loops} \cite{hazelhoff1951structure,biggs1957new,duncker1974structure}. Indeed, the directed flows are observed along looped airways formed by dorso-, para- and ventro-bronchi (DPV), as illustrated in the anatomical schematic of Fig. \ref{fig1}(a). This respiratory network is thus broadly analogous to a circulatory system, with alternating or AC forcing via air sac contraction and expansion driving directional or DC flows around loops \cite{hazelhoff1951structure,brackenbury1971airflow}. However, no valvular structures have been found in bird lungs \cite{duncker1974structure,king1958volumes}, spurring various hypotheses for rectification that invoke: static baffle-like structures that `guide' flows \cite{hazelhoff1951structure}; bronchial tube constrictions and divergences that induce separated flows and jets \cite{banzett1991pressure,wang1992aerodynamic,maina2000inspiratory,sakai2006numerical}; the curvature of tubes and their angles at junctions \cite{brackenbury1972physical, maina2009inspiratory}; and phased expansion/contraction of multiple air sacs and their compliance \cite{urushikubo2013effects,harvey2016robust}. Importantly, the Reynolds numbers $\textrm{Re} \sim 100-1000$ are high  \cite{bretz1971bird,butler1988inspiratory,kuethe1988fluid}, and studies that vary gas density and flow speed indicate the importance of inertial effects \cite{banzett1987inspiratory,kuethe1988fluid}. In particular, studies by Banzett on anesthetized birds point to an inherently aerodynamic source of valving in which inertia or momentum directs flows at junctions  \cite{banzett1987inspiratory,wang1988bird,butler1988inspiratory,banzett1991pressure,wang1992aerodynamic}.

Building on these works, here we test and verify that loopy networks driven with oscillations or pulsations at high Reynolds numbers represent sufficient ingredients for AC-to-DC conversion. Towards isolating the physical factors essential for rectification, we begin with a series of geometrical idealizations of the avian respiratory network. In Fig. \ref{fig1}(b) we show a `spherical bird' approximation that isolates one DPV arc, which connects to a single caudal air sac (AS) on one end and to the primary bronchus (P) on the other, the latter idealized as a tube that opens to the outside at the mouth. A closed loop is formed by DPV and the mesobronchus (M). The air sac functions as a bellows that reciprocally expands and contracts to drive inspiration and expiration. The system is transformed into a closed circuit by replacing the air sac with a reciprocating piston and rerouting the main bronchus to attach as shown in Fig. \ref{fig1}(c). This closure is an experimental and computational convenience for studying flows internal to the network and without regard to exchange with the external environment.

Laboratory demonstrations prove the rectification capability of the circuit of Fig. \ref{fig1}(c), as shown in the Supplemental Video 1. A system is constructed from clear rubber tubing and T-junction fittings, filled with water and flow-tracing particles, and driven via a reciprocating roller that serves the function of a piston. While the flow in the lower portion (yellow and orange) matches the driving motions and is purely AC, the upper segment (light green) displays directed or DC flows. Their sense is indicated by the arrow in Fig. \ref{fig1}(c) and matches that observed in the DPV of bird lungs. Further, DC motions are apparent in the straight segment (dark green, M) connecting the two junctions. In essence, circulation develops around the closed loop that includes DPV and M, and the system acts as a circulating pump, fluidic rectifier or AC-to-DC converter.

The network of Fig. \ref{fig1}(c) contains asymmetries not critical to rectification, as demonstrated by considering further transformations that result in the symmetrized circuit of Fig. \ref{fig1}(d). The lengths and trajectories of the conduits are manipulated while preserving the network topology, junction geometry and connectivity. Specifically, the side branch of each T-junction feeds into the straight segment of the partner junction. This can be achieved in various 3D arrangements and in the quasi-2D layout of Fig. \ref{fig1}(d), which is planar except for a slight defect where the conduits cross. Demonstrations again show persistent DC circulation for AC driving. The network is mirror symmetric and so are the time-averaged DC flows, which emanate from the straight branches and return through the side branches of the junctions, as indicated in Fig. \ref{fig1}(d). These observations rule out an explanation of the observed rectification based on spontaneous symmetry breaking.

\begin{figure*}
\centering
\includegraphics[width=17cm]{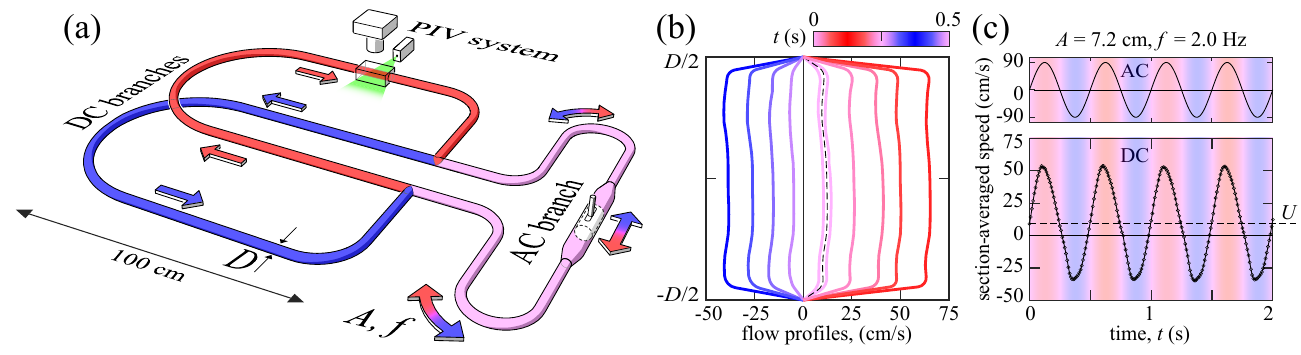}\vspace{-0.2cm}
\caption{Experimental system for characterizing flows in a forced fluidic network. (a) Rigid tubing of inner diameter $D = 1.6$ cm is formed and connected via T-junctions, and the network is filled with water seeded with microparticles. Oscillatory flow in the AC branch is driven by a motorized piston that oscillates with amplitude $A$ and frequency $f$, and the resulting flow field in a DC branch is measured via PIV. (b) Flow profiles in the DC branch measured at different times in the cycle for $A = 7.2$ cm and $f = 2.0$ Hz ($A/D = 4.5$, $\textrm{Wo}^2 = 900$ and $\textrm{Re} = 2500$). Rectification is evident as the positive bias in the time-averaged profile (dashed curve). (c) Imposed oscillatory flow in the AC branch (top) and measured section-averaged flow in a DC branch (bottom); the time-average $U$ of the latter characterizes the pump rate.}
\label{fig2}\vspace{-0.1cm}
\end{figure*}

More careful experiments show that the rectified flows are reproducible for given forcing parameters and vary systematically with these parameters. As shown in the experimental schematic of Fig. \ref{fig2}(a), a version of the symmetric system is constructed from rigid tubing and custom-made junctions (all of uniform inner diameter $D = 1.6$ cm), filled with water, and connected to a motorized piston that oscillates sinusoidally with controllable amplitude $A \sim 0.3-8$ cm and frequency $f \sim 0.3-3$ Hz. This yields high Reynolds numbers that span those relevant to bird respiration \cite{bretz1971bird,butler1988inspiratory,kuethe1988fluid}: $\textrm{Re} = \rho A f D/ \mu \sim 10-5000$, where $\rho = 1.0~\textrm{g}/\textrm{cm}^{3}$ and $\mu = 8.9 \times 10^{-3}~\textrm{dyn} \cdot \textrm{s} / \textrm{cm}^{2}$ are the density and viscosity of water, respectively. Seeding with microparticles allows for flow measurement via particle image velocimetry (PIV). A portion of the pipe in one of the DC branches is illuminated by a laser sheet and imaged with a high-speed camera, and a rectangular water jacket and thin-walled tubing ensure minimal optical distortion. The measured particle motions yield time- and space-resolved flow velocity fields. Profiles of the axially-averaged velocity, sampled throughout an oscillation period, are shown in Fig. \ref{fig2}(b) for selected values of $A$ and $f$. Rectification manifests as a bias towards positive velocities. Assuming that the flow is axisymmetric, the radial average of the flow velocity yields the section-averaged speed shown in Fig. \ref{fig2}(c). Its time average $U$ quantifies the emergent DC flow speed or, equivalently, the time- and section-averaged volumetric flux (volume flow rate per unit area).

\begin{figure*}
\centering
\includegraphics[width=16.5cm]{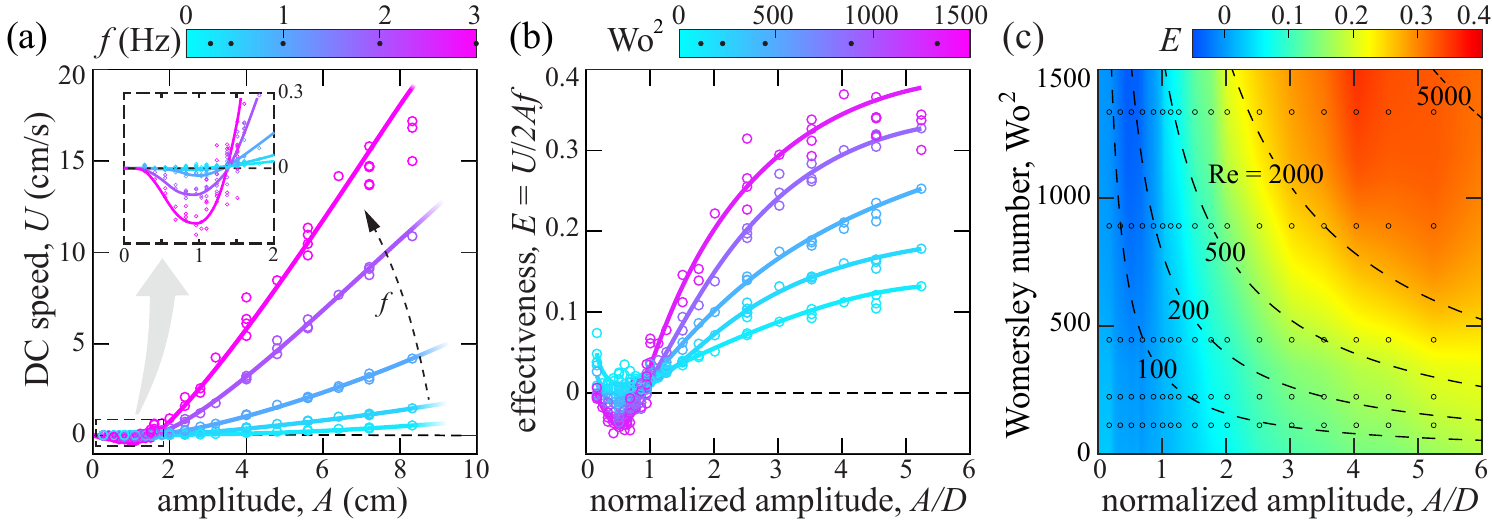}\vspace{-0.2cm}
\caption{Experimental characterization of AC-to-DC conversion in the symmetric circuit of Fig. 2. (a) Dependence of DC flow speed $U$ on AC driving parameters $A$ and $f$. Multiple data points represent independent trials, Bezier curves are shown as guides to the eye, and values of $f$ are indicated on the colorbar. Inset: Magnified view of low-$A$ data. (b) Effectiveness $E = U/2Af$ as a dimensionless measure of AC-to-DC conversion. (c) Map of $E$ across normalized amplitude $A/D$ and frequency $\textrm{Wo}^{2} \propto f$. Markers indicate where measurements are performed, and the map is generated by linear interpolation/extrapolation and smoothing via moving-window averaging.}
\label{fig3}\vspace{-0.4cm}
\end{figure*}

The performance as an AC-to-DC converter can be characterized by measuring the output $U$ across different values for the input parameters $A$ and $f$. The plots of Fig. \ref{fig3}(a) show systematic trends in these data, and repeated measurements at each $A$ and $f$ indicate reproducibility of the phenomenon and reliability of the measurements. The pump rate increases with $A$ and $f$ for all but the smallest values of these parameters, for which $U$ is small or even negative (inset).

A dimensionless measure of pumping effectiveness is $E=U/2Af$, which compares the output DC speed to the characteristic input AC speed. Quantitatively, $E = 1$ represents perfect rectification in the following sense. A stroke of the piston in one direction, say the downward motion marked in red in Fig. \ref{fig2}(a), displaces fluid in the AC branch by an amount $2A$ in the duration $1/2f$. The entire flux $4Af$ is injected straight past the lower T-junction and into the DC branch colored red, and it returns in whole to the other side of the AC branch by turning at the upper junction. The side branch of the lower junction behaves as if sealed shut by a valve, deactivating the other DC branch (blue). The DC branches and T-junctions swap roles in the return stroke. This ideal behavior is thus half-wave rectification with cycle-averaged flux $U=2Af$ in each DC branch, hence $E=1$. Figure \ref{fig3}(b) shows the measured $E(A,f)$ to be as high as 0.4, and the trends indicate increasing $E$ for greater $A$ and $f$.

The normalization $E = U/2Af$ fails to collapse these data, highlighting the nonlinear response of the system. A fully dimensionless characterization can be formed in terms of $A/D$, which represents the oscillatory amplitude relative to conduit diameter, and the square of the Womersley number $\textrm{Wo}^{2} =  \pi \rho f D^2 / 2\mu$, which is proportional to $f$ and assesses the unsteadiness of pulsatile flow \cite{womersley1955method}. Figure \ref{fig3}(c) displays $E(A/D,\textrm{Wo}^{2})$, showing increasingly effective pumping in both driving parameters. Also shown are the hyperbolic contours of the Reynolds number $\textrm{Re} = (2/\pi) (A/D) \textrm{Wo}^{2}$.

\begin{figure*}
\centering
\includegraphics[width=17cm]{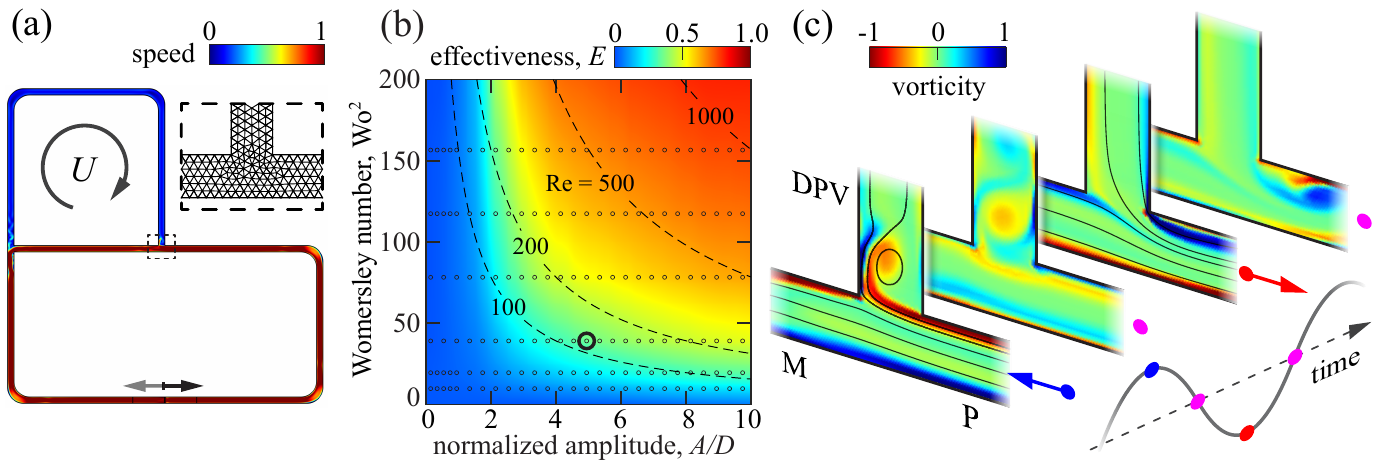}\vspace{-0.2cm}
\caption{Simulation of the asymmetric circuit. (a) The domain and spatial grid near a T-junction (inset). Color indicates flow speed normalized by $2 \pi fA $ and at the moment the piston is moving rightward at peak velocity. Here $D = 0.5$ cm, $A = 2.5$ cm and $f = 1$ Hz. (b) Effectiveness $E$ across $A/D$ and $\textrm{Wo}^2 \propto f$. (c) Streamlines and vorticity near one junction at four equally spaced phases in the cycle and for the parameters circled in (b). Arrows indicate the phase of the imposed flow from the MB. Vorticity is normalized by $ \pi^2f(A/D)^2 $, a scale appropriate for vortex shedding from a sharp edge in oscillatory flow \cite{schnipper2009vortex,agre2016linear}.}
\label{fig4}\vspace{-0.4cm}
\end{figure*}

To further validate these findings, we implement computational fluid dynamics simulations relevant to the asymmetric circuit of Fig. \ref{fig1}(c). The two-dimensional Navier-Stokes equation for incompressible flow is numerically solved using the finite element method in the domain defined in Fig. \ref{fig4}(a), where no-slip boundary conditions are imposed on the static walls as well as the reciprocating piston, the neighborhood of which is treated using an Arbitrary Lagrangian-Eulerian (ALE) formulation \cite{hirt1974arbitrary}. The simulations are carried out using the COMSOL Multiphysics software package with laminar flow settings \cite{comsol}. A detail of the spatial grid near one T-junction is shown in the inset. The system parameters span the ranges $A/D \sim 0.1-10$, $\textrm{Wo}^{2} \sim 10-200$ and $\textrm{Re} \sim 10-1000$. Compared to our experiments, the simulations are restricted to lower $\textrm{Wo}^{2}$ but permit larger $A/D$, leading to comparable Re. Additional implementation details are available as Supplemental Material.

We assess the dependence of the terminal or long-time mean flow rate $U$ on the driving parameters $A$ and $f$. Our results reproduce all phenomenology and trends seen in experiments, as shown by the $E(A/D,\textrm{Wo}^{2})$ map in Fig. \ref{fig4}(b) and the associated curves in the Supplemental Material. The simulations typically yield values of $E$ that are several times greater than those measured at corresponding conditions in experiments. This may be attributed to dimensionality (2D simulations versus 3D experiments), the use of asymmetric versus symmetric forms of the circuit, and the consequent differences in the lengths and curvatures of the conduits. These factors affect the resistance of DPV and M, and the longer conduits used in experiments and thus higher resistance may contribute to the lower effectiveness. Nonetheless, the features of emergent DC flows, their sense of direction, and qualitative trends for varying driving parameters are robust to these geometric differences.The maps of Figs. \ref{fig3}(c) and \ref{fig4}(b) indicate that, for large $\textrm{Re}$, contours of constant $\textrm{Re}$ roughly correspond to constant $E$, suggesting that extrapolation may be used to estimate $E$ for values of $A/D$ and $\textrm{Wo}^{2}$ larger than those explored here.

The simulations provide access to the flow fields, as shown in Supplemental Videos 2 and 3, and thereby offer insights into the rectification mechanism. In Fig. \ref{fig4}(c) we focus on streamlines and vorticity near the junction highlighted in (a) and at four instances in the oscillation cycle. At peak velocity during inhalation (first image), fluid is injected from P and predominantly goes straight past the junction down M, with little turning up DPV. This matches the intuition that inertia of the flow tends to maintain its straight course. A stagnation streamline impinges on the far (left) corner of the junction and divides the straight and turning flows, and a large vortex shed from the near (right) corner `plugs' the side branch. At peak velocity during exhalation (third image), fluid exits the junction via P and draws more equally from M and DPV. The converging currents are divided by a separation streamline emanating from the far corner, and vorticity produced at the near corner does not detach but rather `hugs' the wall of P. Integrating in time over a cycle, the indicated junction thus has net flux incoming from the side branch (DPV) and exiting via the straight branch (M), which corresponds to net circulation in the clockwise sense around the DPV-M loop. Consistent with mass conservation, the other junction also has net flux entering via its side branch and exiting via the straight branch, these associated with M and DPV, respectively. In essence, flow separation and vortex shedding serve the valving function of closing and opening the side branch with appropriate timing in the cycle. Analogous phenomena are expected to occur in 3D settings, which could be assessed through flow visualization and quantification in experiments as well as 3D simulations.

These results complement the physiological studies by Banzett and others on the aerodynamic origins of valving in birds during inspiration \cite{banzett1987inspiratory,wang1988bird,butler1988inspiratory,banzett1991pressure,wang1992aerodynamic} by providing realizations of driven networks in physical experiments and simulations. We have demonstrated the unsteady and separated nature of the causal flows at junctions and characterized the consequent nonlinear response over relevant fluid dynamical parameters. Taken together, these results establish a form of fluidic rectification that is ultimately rooted in kinematic irreversibility at high Reynolds numbers \cite{tritton2012physical,schlichting2016boundary}, which plays an analogous role in contexts such as expulsion versus suction from orifices and reverse versus forward flow through asymmetric conduits. The relevant asymmetry in our system is more subtle: Each T-junction has an axis of symmetry, but its anisotropic shape has distinct straight and side branches with differing end conditions due to their connectivity to the partner junction. Rectified flows of the type observed here are impossible for $\textrm{Re}=0$ or viscous-dominated conditions, for which the governing Stokes equation is reversible and the Scallop Theorem ensures flows induced in one stroke are retraced reversely in the return stroke \cite{tritton2012physical,purcell1977life}. Rectification is also precluded in the other extreme of inviscid and irrotational flow due to Kelvin's Circulation Theorem \cite{tritton2012physical}. Hence, the valving mechanism described here may operate at any finite Re, though it is expected to be exceedingly weak for $\textrm{Re} < 10$. The turbulent flow regime of $\textrm{Re} > 2000$, while less relevant to bird lungs, may be applicable to other problems and awaits future studies.

The phenomena reported here may arise more generally in the looped topologies common across many biological and physiological flow networks, which are often subject to unsteady forcing and pulsatile flows \cite{schmidt1997animal,fung2013biomechanics-circulation}. Loops play the essential role of providing routes around which circulation can be established. Our systems are minimal in that they comprise two interconnected loops, one whose AC flow is prescribed and the other free to display an emergent DC flow. Loops necessitate junctions, and the mechanism described above suggests that their geometrical anisotropy is critical. We posit that any anisotropy, appropriately mirrored for partner junctions, will in general induce rectification. In this sense, the T-shape used here is not critical, and the complexities of junction flows invite optimization over parameters such as branching angles \cite{wang1988bird,aultandstone2016vortex,takagi1985study,haselton1982flow,mauroy2003interplay}. Future studies might vary not only the junction and conduit geometries but also the driving waveform and global network topology and connectivity, which could further inform on respiration phenomena and flow transport in complex networks generally. The value of the simpler circuits studied here is in identifying loops, anisotropic junctions and inertial flows as sufficient ingredients for rectification.

\textit{Acknowledgements.} The authors thank C. Peskin, M. Shelley and J. Zhang for useful discussions and acknowledge support from the NSF (DMS-1720306 to C.F., DMS-1646339 and DMS-1847955 to L.R.) and the Simons Foundation (Collaboration Grant for Mathematicians, Award 587006 to A.U.O.). 

\bibliography{apssamp}

\end{document}